\documentclass[aps,preprint,llncs]{revtex4}
\usepackage[utf8]{inputenc}
\usepackage[english]{babel}
\usepackage[T1]{fontenc}
\usepackage{amsmath}
\usepackage{amsfonts}
\usepackage{ragged2e}
\usepackage{amssymb}
\usepackage{xcolor}
\usepackage[colorlinks=true,linkcolor=blue,filecolor=blue, urlcolor=blue,citecolor=blue]{hyperref}
\usepackage{graphicx}
\usepackage{float}
\usepackage{color}
\usepackage{soul}
\usepackage{siunitx}
\usepackage{physics}
\usepackage{lineno}
\usepackage{subfiles}

\begin{document}

\title{Nucleation and Enhancement of Superconductivity under Tip-Induced Strain Fields}

\author{Ghulam Mohmad, Vishal Tripathi and Goutam Sheet}
\email{goutam@iisermohali.ac.in}
\affiliation{Department of Physical Sciences, Indian Institute of Science Education and Research (IISER) Mohali, India}

\begin{abstract}
A metallic point contact formed on a non-superconducting or weakly 
superconducting material often nucleates or enhances superconductivity confined under the contact. However, no unified theoretical description of the phenomenon exists. We show  that the spatially inhomogeneous, predominantly uniaxial nature of the stress field under a point contact is fundamental for such tip-induced and  tip-enhanced superconductivity (TISC/TESC). We also show that the coupling of such a stress field to the 
electronic structure can be estimated through an experimentally measurable 
uniaxial coupling scale $C^{\mathrm{exp}}$. Combining Hertzian contact 
mechanics with a Ginzburg-Landau variational analysis, we derive a 
criterion for the nucleation of TISC/TESC and determine $C^{\mathrm{exp}}$ for 
twenty-one materials. For topological semimetals with ungapped band crossings, 
the framework explains observed critical temperatures with no free 
parameters and for all others, $C^{\mathrm{exp}}$ provides a direct
experimental determination of the uniaxial strain sensitivity and a 
target scale for microscopic theories.The work predicts TISC in elemental Sb and Y with $T_c \approx 2.8$\,K and $T_c \approx 12$\,K respectively.
\end{abstract}

\maketitle

It is known that when a metallic tip is pressed against a non-superconducting
material, superconductivity sometimes appears exclusively at the contact.  The same
geometry also often raises the local $T_c$ above its bulk value in a material that
is already superconducting.  These phenomena, tip-induced superconductivity (TISC)
and tip-enhanced superconductivity (TESC) respectively, share common experimental signatures, namely conductance enhancement via Andreev reflection in the ballistic regime, or
critical-current-driven dips in the thermal
regime~\cite{Aggarwal2016NatMat,Wang2016NatMat, Wang2018_review,Howlader2021_review}.  TISC has been reportedd in Cd$_3$As$_2$, TaAs,
WC, NbAs$_2$, grey~As, Pb$_{0.6}$Sn$_{0.4}$Te, ZrSiS, and other topological
semimetals~\cite{Aggarwal2016NatMat,Wang2016NatMat,Aggarwal2017NatComm,Hou2019PRB,%
Zhang2020PRB,Das2016APL,GreyAs2023,Aggarwal2019ZrSiS}, whereas TESC appeared in TaAs$_2$, Sr$_2$RuO$_4$, Au$_2$Pb, and
several transition-metal
dipnictides~\cite{Zhang2020PRB,Hou2019PRB,Wang2015PRB_SRO,Xing2016Au2Pb}.  Despite
growing experimental evidence for the phenomena in wider variety of quantum materials, a theoretical understanding has been lacking.

The most direct consequence of a tip pressing on a sample in a usual point contact geometry is a large spatially
inhomogeneous stress field that locally modifies lattice parameters, phonon spectra,
and electronic bandwidth.  Traditionally, the observed $T_c$ shift under such point contact environments has been compared with bulk hydrostatic
pressure coefficients~\cite{Wang2018_review,Howlader2021_review,Zhang2021CPB}. This misses
two important features.  First, the stress is localized and decays over the contact radius $a$. Under this condition the question is
not merely whether the developed strain shifts $T_c$, but whether a self-sustained superconducting pocket can be nucleated against the confinement cost of
the order parameter.  Ginzburg-Landau (GL) theory near flat interfaces addresses
confinement~\cite{deGennes1964,Fink1969} but treats only uniform geometries and
does not give a load-dependent threshold.  Second, the stress under a point contact is
predominantly uniaxial, and uniaxial strain can couple differently. For example, uniaxial strain couples to ungapped Dirac and Weyl band
crossings far more strongly than isotropic compression in certain topological semimetals. We emphasize that these important features are fundamentally important for understanding TISC/TESC.

To understand the possible distribution of stress under a point contact, we first refer to Hertzian theory of contact mechanics~\cite{Johnson1985,Popov2019,Zhu2012}. When a spherical metallic tip of radius $R$ presses onto a flat surface under
normal load $F$, Hertzian mechanics gives a contact radius $a$  over which the non-uniform stress exists, and the maximum pressure ($p_{\max}$):
\begin{equation}
a = \left(\frac{3FR}{4E^*}\right)^{1/3},  \qquad p_{\max} = \frac{3F}{2\pi a^2},
\end{equation}
where $1/E^* = (1-\nu_t^2)/E_t + (1-\nu_s^2)/E_s$, with $E_{t,s}$, $\nu_{t,s}$
the Young's moduli and Poisson ratios of tip and sample.  The full subsurface stress
field is obtained from superposing Boussinesq point-load
solutions~\cite{Johnson1985,Popov2019}.  Near the contact axis, the axial normal stress component of the subsurface stress tensor $\sigma_{zz}$
dominates, making the field predominantly uniaxial for $r\ll a$.  The peak axial
strain at ($r=0$, $z=a$) is
\begin{equation}
\varepsilon_{zz}^{\rm peak} \approx \frac{0.96\,p_{\max}}{K},
\label{eq:eps_zz}
\end{equation}
for $\nu_s\approx0.25$ (representative of materials under discussion here), where $K=E_s/[3(1-2\nu_s)]$ is the bulk modulus.

\begin{figure}
\centering
\includegraphics[width=0.9\columnwidth]{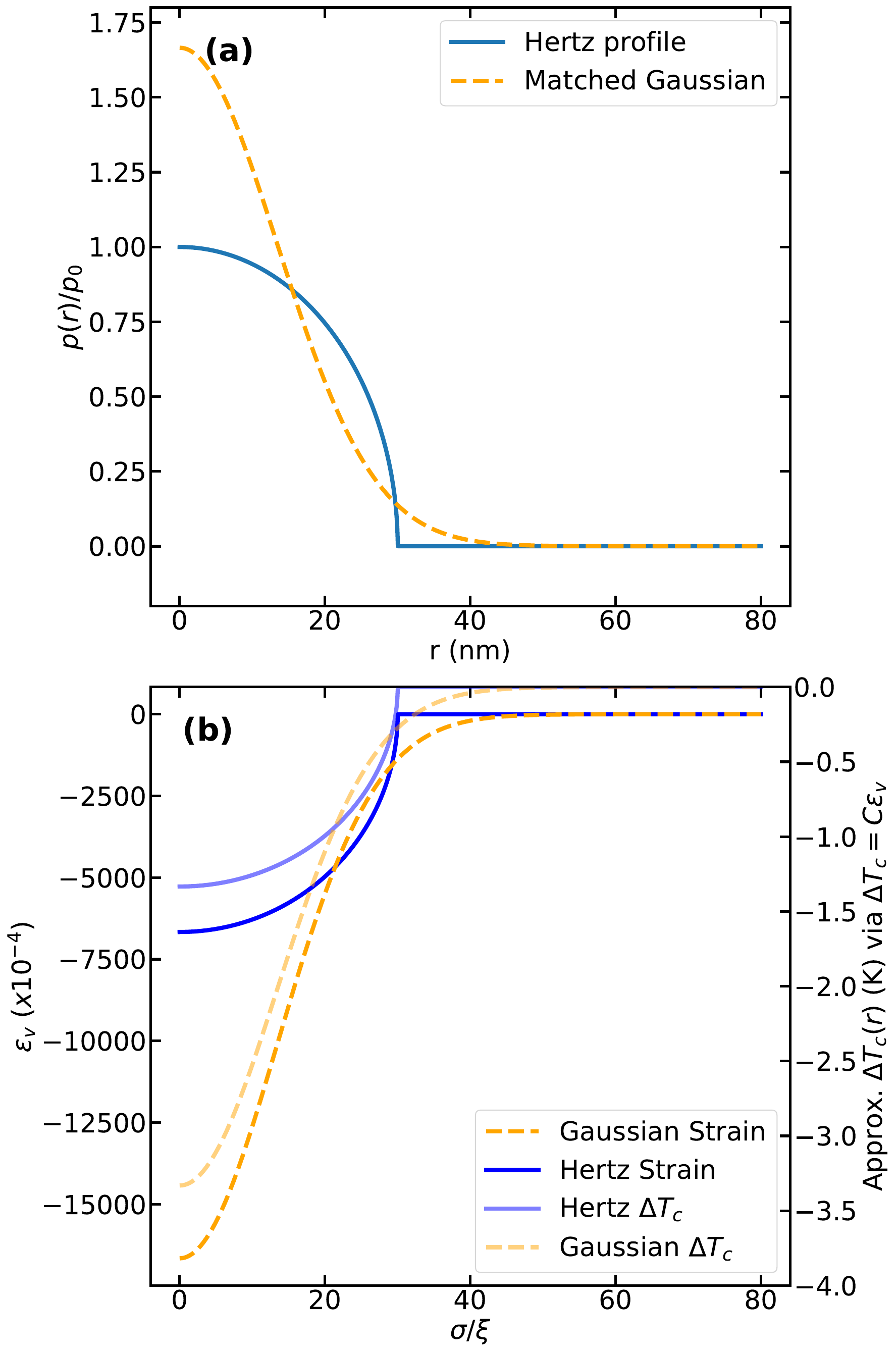}
\caption{(a)~Hertzian pressure profile and matched Gaussian ($\sigma\approx0.447\,a$).
(b)~Axial strain $\varepsilon_{zz}(r)$ and $\Delta T_c(r)$ for both profiles;
agreement is within 5\% everywhere.}
\label{fig:strain_and_Q}
\end{figure}

For analytical simplicity we replace the Hertz profile by a matched Gaussian,
\begin{equation}
p_{\mathrm{eff}}(\mathbf{r})=p_{\max}\exp\!\left[-\frac{r^2+z^2}{2\sigma^2}\right],
\label{eq:p}
\end{equation}
with $\sigma = a/\sqrt{5}$ fixed by matching the second moment of the Hertz surface
pressure distribution~\cite{Zhu2012}. As
shown in Fig.~\ref{fig:strain_and_Q}, this replacement leads to a very small error $\sim$ below 5\% everywhere.  Now, the local $T_c$ shift is assumed to follow linear response to strain and the local strain is assumed to be related to the local pressure through Hooke's law: 
\begin{equation}
\Delta T_c(\mathbf{r}) = C\,\varepsilon(\mathbf{r}) \approx \frac{C}{K}\,p_{\mathrm{eff}}(\mathbf{r}),
\label{eq:dTc}
\end{equation}
where $C \equiv dT_c/d\varepsilon > 0$ is the strain-sensitivity
coefficient~\cite{DeGennes1999}. This coefficient captures whatever microscopic mechanism
couples the generated strain field to superconducting pairing e.g., phonon softening, band-structure modification, or charge redistribution.  The maximum $T_c$ shift occurs at the contact centre where $p_{\rm eff}$ is
largest, giving $\Delta T_c^{\rm th} \approx (C_{\rm th}/K)\,p_{\max}$.  When the contact $T_c$ is known from a TISC/TESC experiment, the observed transition
temperature $T_c^{\rm obs}$ corresponds to the point of maximum strain at the
contact centre, since superconductivity persists until the last and most strongly
coupled region of the junction becomes normal. Thus, Eq.~\eqref{eq:dTc}
gives $C^{\rm exp} \approx \Delta T_c^{\rm obs} \cdot K/p_{\max}$. This parameter directly provides the scale of the coupling strength between the inhomogeneous uniaxial
strain field and the electronic structure of the quantum material under investigation.

\begin{figure}[t]
\centering
\includegraphics[width=0.9\columnwidth]{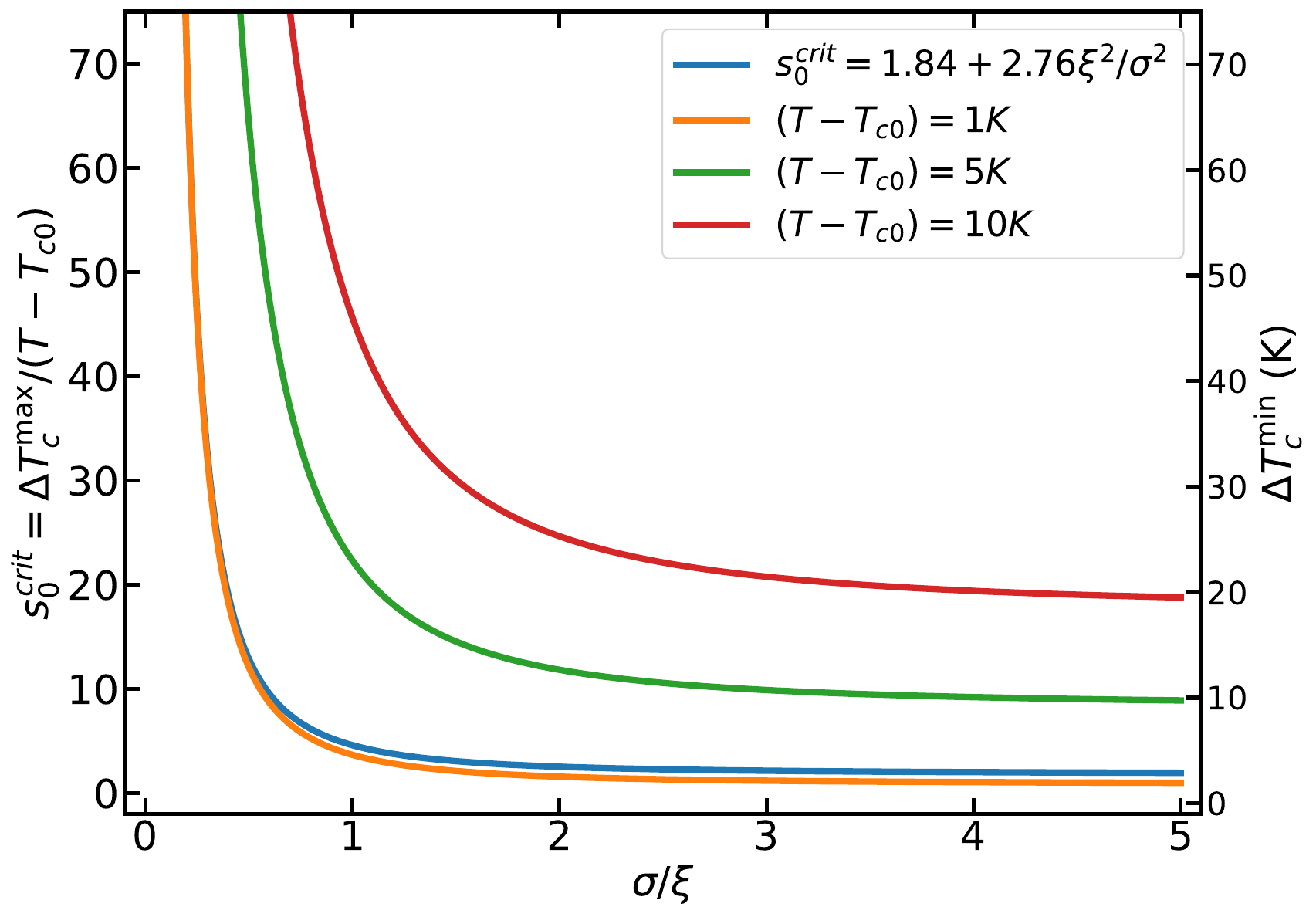}
\caption{Nucleation threshold $s_0^{\rm crit}=\Delta T_c^{\max}/(T-T_{c0})$ vs.\
$\sigma/\xi$ (Eq.~\eqref{eq:dTcmin}); other lines show $\Delta T_c^{\max}$ for
representative $T-T_{c0}$.}
\label{fig:nucleation_threshold}
\end{figure}

The spatially varying $T_c(\mathbf{r})$ suggests that the local instability
toward superconducting order is strongest at the contact centre and decays
away from it.  Whether this spatially confined tendency is sufficient to
nucleate a self-sustained condensate depends on the competition between the
gain in condensation energy at the contact centre and the kinetic cost of
confining the order parameter to a small region.  Resolving this competition
requires a free-energy analysis, for which we employ GL theory.  The order
parameter $\psi(\mathbf{r})$ is governed by the free-energy functional
\begin{equation}
F[\psi] = \int d^3r \left[ \alpha(\mathbf{r},T)|\psi|^2 +
\frac{\beta_{\rm GL}}{2}|\psi|^4 + \xi_0^2|\nabla\psi|^2 \right],
\label{eq:GL_free}
\end{equation}
where $\alpha(\mathbf{r},T) = \alpha_0(T - T_c(\mathbf{r}))$, with $\alpha_0
> 0$ a material-dependent GL coefficient, $\beta_{\rm GL} > 0$ the nonlinear
coefficient, and $\xi_0$ the bare coherence length.  Here $T_c(\mathbf{r}) =
T_{c0} + \Delta T_c(\mathbf{r})$, where $T_{c0}$ is the bulk transition
temperature of the host material: $T_{c0} = 0$ for TISC (the host is not
superconducting) and $T_{c0} > 0$ for TESC.  Near onset $|\psi|$ is small
and the nonlinear term can be neglected.  Setting $\xi =
\xi_0/\sqrt{\alpha_0(T - T_{c0})}$ as the GL coherence length at temperature
$T$, and $s(\mathbf{r}) = \Delta T_c(\mathbf{r})/(T - T_{c0})$:
\begin{figure}[H]
\centering
\includegraphics[width=0.9\columnwidth]{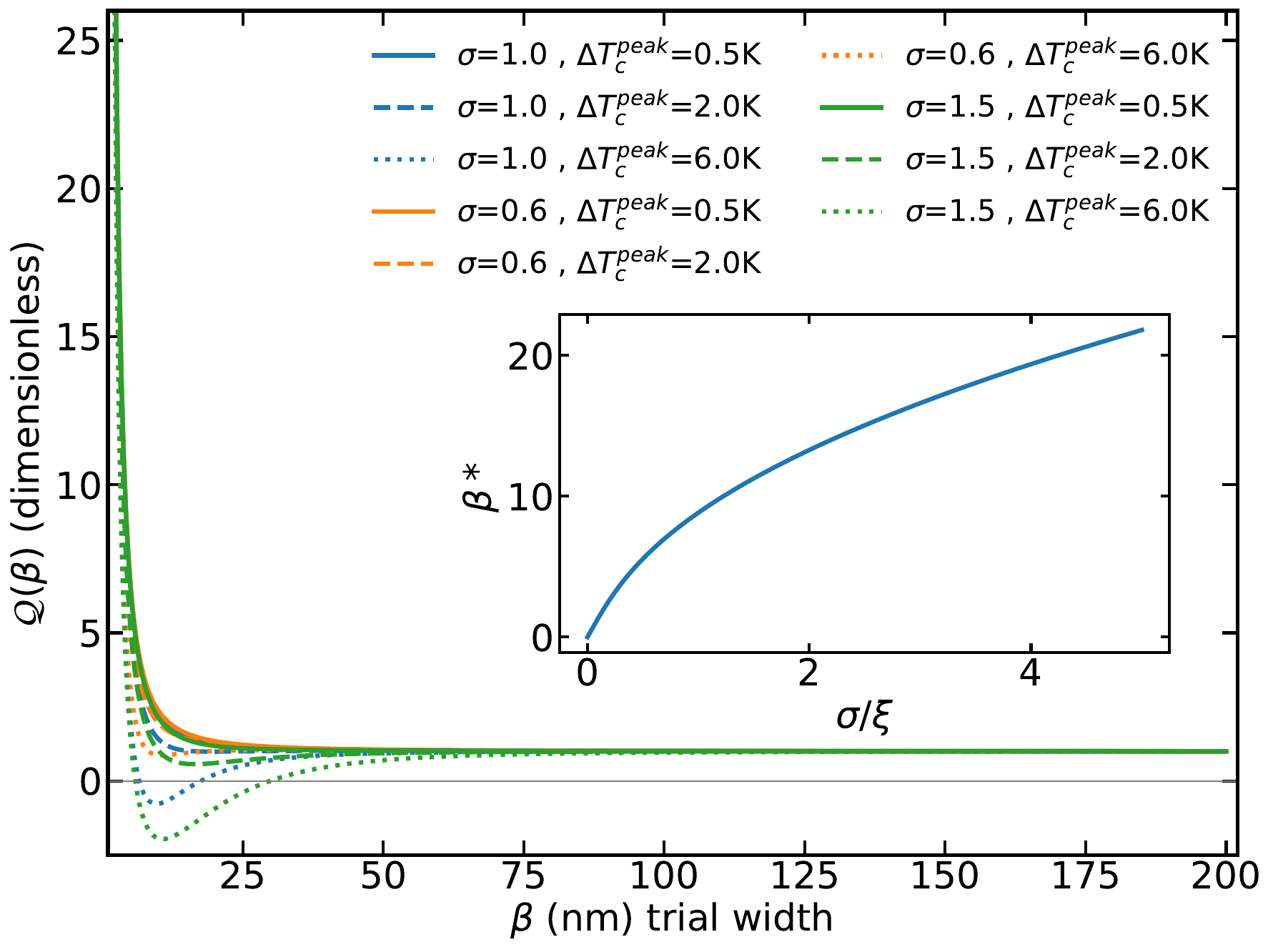}
\caption{Rayleigh-Ritz functional $\mathcal{Q}(\beta)$ for several $\sigma/\xi$
ratios; nucleation sets in when $\mathcal{Q}=0$.  Inset: optimal width $\beta^*$
vs.\ $\sigma/\xi$.}
\label{fig:hertz_gaussian}
\end{figure}

\begin{equation}
-\xi^2\nabla^2\psi + [1-s(\mathbf{r})]\psi = 0
\label{eq:scaled}
\end{equation}
This is a Schr\"{o}dinger-like equation with an attractive potential well
$-s(\mathbf{r})$ centred at the contact, whose depth is set by the magnitude
of the local $T_c$ enhancement and whose width is set by $\sigma$.  Since
$s(\mathbf{r}) \propto p_{\rm eff}(\mathbf{r})$ as evident from Eqs.~\eqref{eq:dTc}
and~\eqref{eq:p}, the potential well is Gaussian. By direct analogy with quantum mechanics, a self-sustained superconducting
pocket exists if and only if this potential well supports a normalizable
bound state, i.e., the local $T_c$ enhancement must be strong enough and wide
enough to bind the order parameter against the kinetic cost of confinement.
The nucleation threshold is the vanishing of the lowest eigenvalue of
$-\xi^2\nabla^2 + [1-s(\mathbf{r})]$, which has no closed-form solution
for a Gaussian potential. We therefore estimate it variationally where the
Rayleigh-Ritz quotient is
\begin{equation}
\mathcal{Q}[\psi]=
\frac{\int d^3r\,[\xi^2|\nabla\psi|^2+(1-s(\mathbf{r}))|\psi|^2]}
{\int d^3r\,|\psi|^2},
\label{eq:Q}
\end{equation}
with nucleation possible when $\min_\psi\mathcal{Q}=0$. Given the Gaussian symmetry
of the potential, we take the trial function $\psi(\mathbf{r}) =
\exp[-(r^2+z^2)/(2\beta^2)]$, where $\beta$ is a variational width parameter
representing the spatial extent of the superconducting pocket. Note that
$\beta \neq \sigma$ in general because the coherence length $\xi$ penalises sharp gradients and
forces the condensate to spread beyond the stress-enhanced region. This leads to (Fig.~\ref{fig:hertz_gaussian})
\begin{equation}
\mathcal{Q}(\beta)=\frac{3}{2}\frac{\xi^2}{\beta^2}
+1-s_0\!\left(\frac{2}{(\beta/\sigma)^2+2}\right)^{3/2},
\label{eq:Qbeta}
\end{equation}
where $s_0=\Delta T_c^{\max}/(T-T_{c0})$.  Minimising over $\beta$ and imposing
$\mathcal{Q}(\beta^*)=0$ yields
\begin{equation}
\frac{3\xi^2}{\beta^{*2}} = s_0\,\frac{6\sqrt2(\beta^*/\sigma)^2}{\bigl[(\beta^*/\sigma)^2+2\bigr]^{5/2}},
\label{eq:betastar}
\end{equation}
and the nucleation threshold
\begin{equation}
\Delta T_c^{\max}\gtrsim\!\left(1.84 + 2.76\,\frac{\xi^2}{\sigma^2}\right)(T-T_{c0}).
\label{eq:dTcmin}
\end{equation}
The coefficients are accurate to better than 2\% for $0.5\leq\sigma/\xi\leq5$.
An anisotropic trial function lowers $s_0^{\rm crit}$ by at most 8\%, so
Eq.~\eqref{eq:dTcmin} is a conservative upper bound
(Fig.~\ref{fig:nucleation_threshold}). The criterion suggests that the coupling between the
tip-induced uniaxial stress and the electronic structure of the material must
be strong enough so that the threshold is crossed in order for TISC/TESC to be nucleated.

We now apply this criterion to the twenty TISC/TESC systems listed in
Table~\ref{tab:TISC_summary}.  The most important material-specific quantity is the
coupling coefficient $C$, which determines $\Delta T_c^{\rm max}$ through
Eq.~\eqref{eq:dTc}. For a meaningful classification of the materials we first define various coupling scales used in the table. $C_{\rm uni} \equiv dT_c/d\varepsilon_{zz}$
is the uniaxial strain sensitivity.  $C_{\rm vol} \equiv K\,dT_c/dP$ is the
volumetric sensitivity extracted from hydrostatic pressure experiments. This
gives a conservative lower bound on $C_{\rm uni}$.
$C_{\rm th}$ is the best available theoretical estimate of $C_{\rm uni}$. For Class~I it is computed directly via the McMillan formula from the
strain-dependence of the node separation; for Classes~II and~III, where no
such computation is available, $C_{\rm th} = C_{\rm vol}$ is used as that
lower bound.  Finally, $C^{\rm exp} \approx \Delta T_c^{\rm obs} \cdot
K/p_{\max}$ is the uniaxial coupling scale inferred by inverting
Eq.~\eqref{eq:dTc} against the experimentally observed $T_c$. $C^{\rm exp}$ is the
quantity that a microscopic theory explaining TISC/TESC must reproduce.

\begin{table*}
\centering
\footnotesize
\setlength{\tabcolsep}{3pt}
\begin{tabular}{l c c c c c c c c}
\hline\hline
\textbf{Material} & \textbf{Bulk $T_c$} & \textbf{Contact $T_c$} & \textbf{$p_{\max}$} & \textbf{$K$} & \textbf{$C_{\rm th}$} & \textbf{$\Delta T_c^{\rm th}$} & \textbf{$\Delta T_c^{\rm obs}$} & \textbf{$C^{\rm exp}$} \\
 & \textbf{(K)} & \textbf{(K)} & \textbf{(GPa)} & \textbf{(GPa)} & \textbf{(K)} & \textbf{(K)} & \textbf{(K)} & \textbf{(K)} \\
\hline
\multicolumn{9}{l}{\textit{Class I: $C^{\rm exp} \approx C_{\rm th} = C_{\rm uni}$ (computed); uniaxial coupling fully accounts for large $C^{\rm exp}$}} \\
\hline
Cd$_3$As$_2$~\cite{Aggarwal2016NatMat} & none & $\sim$7-8 & 2-4 & 100 & $250\pm75$ & $\sim$5-10 & $\sim$8 & $\sim$267 \\
TaAs~\cite{Aggarwal2017NatComm} & none & $\sim$5 & 2-4 & 135 & $270\pm80$ & $\sim$4-8 & $\sim$6 & $\sim$270 \\
Pb$_{0.6}$Sn$_{0.4}$Te~\cite{Das2016APL,KimTCI2017} & none & $\sim$5 & 1-3 & 55 & $140\pm35$ & $\sim$2.5-7.6 & $\sim$5 & $\sim$138 \\
\hline
\multicolumn{9}{l}{\textit{Class II: $C^{\rm exp} \gg C_{\rm vol}$; $C_{\rm uni}$ far exceeds hydrostatic bound; mechanism of large $C_{\rm uni}$ independently identified}} \\
\hline
PdSb~\cite{PdSb2022} & 1.32 & $\sim$3-4 & 1-3 & 90 & 1 & $\lesssim$0.02 & $\sim$2 & $\sim$90 \\
Sr$_2$RuO$_4$~\cite{Wang2015PRB_SRO,PaglioneSRO2002,Steppke2017Science} & 1.5 & $\sim$3.5 & 0.5-2 & 150 & 15 & $\sim$0.03-0.12 & $\sim$2 & $\sim$150 \\
Si~\cite{Sirohi2021} & none & $\sim$10 & few & 99 & - & $\ll$1 & $\sim$10 & $\sim$330 \\
Ge~\cite{Germanium} & none & $\sim$6 & few & 75 & - & $\ll$1 & $\sim$6 & $\sim$150 \\
\hline
\multicolumn{9}{l}{\textit{Class III: $C^{\rm exp} \gg C_{\rm vol}$; $C_{\rm uni}$ far exceeds hydrostatic bound; $C^{\rm exp}$ is the first measurement of $C_{\rm uni}$}} \\
\hline
TaAs$_2$~\cite{Zhang2020PRB,HighPTaAs2016} & none & $\sim$3 & 1-3 & 90 & 10 & $\sim$0.06-0.2 & $\sim$3 & $\sim$135 \\
NbAs$_2$~\cite{Zhang2020PRB} & none & $\sim$3.5 & 1-3 & 85 & 10 & $\sim$0.07-0.2 & $\sim$3.5 & $\sim$149 \\
WC~\cite{Hou2019PRB} & none & $\sim$12 & 2-4 & 380 & 8 & $\sim$0.02-0.05 & $\sim$12 & $\sim$1520 \\
Grey As~\cite{GreyAs2023} & none & $\sim$9 & 2-4 & 22 & 5 & $\sim$0.3-0.5 & $\sim$9 & $\sim$66 \\
ZrSiS~\cite{Aggarwal2019ZrSiS} & none & $\sim$7.5 & 1-3 & 95 & $\lesssim$1 & $\lesssim$0.03 & $\sim$7.5 & $\sim$356 \\
PtBi$_2$~\cite{PtBi2024} & none & $\sim$5 & 2-4 & 90 & 15 & $\sim$0.2-0.4 & $\sim$5 & $\sim$150 \\
TbMn$_6$Sn$_6$~\cite{TbMn6Sn62023} & none & $\sim$2-3 & 1-3 & 120 & 10 & $\sim$0.05-0.15 & $\sim$2-3 & $\sim$150 \\
Zr~\cite{Sirohi2018APL} & 0.57 & $\sim$3.0 & 1-3 & 94 & 3.3 & $\sim$0.02-0.06 & $\sim$2.4 & $\sim$113 \\
Ti~\cite{TipEnhancedTi2024} & 0.40 & $\sim$3.5 & 0.5-3 & 110 & 7.7 & $\sim$0.02-0.12 & $\sim$3.1 & $\sim$227 \\
Ce~\cite{CePointContact,ZhangCe2023,AllenMartin1982} & none & $\sim$20 & 1-3 & 50 & 10 & $\lesssim$0.4 & $\sim$20 & $\sim$500 \\
Au$_2$Pb~\cite{Xing2016Au2Pb,Au2PbElastic2025} & 1.3 & $\sim$2.1 & 0.5-2 & 28 & 5 & $\sim$0.05-0.2 & $\sim$0.8 & $\sim$18 \\
Sb~\cite{OurPRBref} & none & $\sim$2 & 1-3 & 42 & $\lesssim$1 & $\ll$1 & $\sim$2 & $\sim$56 \\
Y~\cite{OurPRBref} & none & $\sim$1.5-12 & 1-3 & 41 & $\lesssim$4 & $\lesssim$0.3 & $\sim$1.5- 12& $\sim$160-600 \\
\hline\hline
\end{tabular}
\caption{Classification of twenty TISC/TESC systems.  $C_{\rm th}$: for Class~I,
computed via McMillan formula from the independently known strain-dependence of the node
separation. For Classes~II--III, $C_{\rm th}=C_{\rm vol}\equiv K\,dT_c/dP$ is a hydrostatic
lower bound only; $C^{\rm exp}\approx\Delta T_c^{\rm obs}\cdot K/p_{\max}$ is then
the first experimental determination of the true uniaxial coupling scale, and the
quantity that microscopic theories of pairing in these materials must reproduce.
The Sb and Y rows were predicted before measurement; see text.}
\label{tab:TISC_summary}
\end{table*}

\textbf{Class I} ($C^{\rm exp} \approx C_{\rm th}$): In a 3D Dirac or Weyl
semimetal, the band-crossing nodes are pinned in momentum space by crystal
symmetry.  Uniaxial strain $\varepsilon_{zz}$ breaks this symmetry
anisotropically and shifts the node separation by $\delta k_D \propto
\varepsilon_{zz}$.  Because the density of states near a Dirac node vanishes
as $N(E) \propto (E - E_{\rm node})^2$, even a modest shift produces a large
fractional change in $N(E_F)$, which through the McMillan relation translates
into a large $\Delta T_c$.  Under hydrostatic pressure, isotropic compression
preserves rotational symmetry and leaves the node separation unchanged; only
bandwidth broadening occurs, suppressing $N(E_F)$~\cite{PressureCd3As2}.
Hence $C_{\rm uni} \gg C_{\rm vol}$ for these materials, and $C_{\rm th}$
can be estimated directly from the strain-dependence of the node separation
via the McMillan formula.  The agreement $C^{\rm exp} \approx C_{\rm th}$
for all three Class~I materials, with no free parameters, constitutes a
direct experimental confirmation of the node-separation mechanism.

For Cd$_3$As$_2$, studies in strained films show a near-fourfold enhancement of
the Dirac node separation~\cite{Pokharel2022PRB}. With
$\lambda\approx0.35\pm0.05$ and
$d(\Delta k_D)/d\varepsilon_{zz}\approx2.0\pm0.4$\,\AA$^{-1}$,
the McMillan formula gives $C_{\rm th}\approx250\pm75$\,K and
$\Delta T_c^{\rm th}\approx5$--10\,K. This is comparable to the observed
$T_c\sim$8\,K~\cite{Aggarwal2016NatMat}. To note, achieving the same effect
hydrostatically requires 8.5\,GPa and a structural
transition~\cite{PressureCd3As2}.  For TaAs, uniaxial strain generates new
Weyl nodes at 1-3\% strain~\cite{TaAsStrain2021,Pandey2024}, giving
$C_{\rm th}\approx270\pm80$\,K ($K\simeq135$\,GPa~\cite{BuckeridgeTaAs2016})
and $\Delta T_c^{\rm th}\approx4$--8\,K. This also matches well with
$T_c \sim$5--6\,K measured experimentally~\cite{Aggarwal2017NatComm}.  For Pb$_{0.6}$Sn$_{0.4}$Te near
the TCI phase boundary~\cite{FuTCI2011,XuTCI2012}, uniaxial strain lifts the
mirror-protected degeneracy; $C_{\rm th}\approx140\pm35$\,K
($K\simeq55$\,GPa) and $\Delta T_c^{\rm th}\approx2.5$--7.6\,K against
$\sim$5\,K from experiments~\cite{Das2016APL}.  All three remain Class~I even at the
$1\sigma$ lower bound of $C_{\rm th}$.  It is important to note that the load-scaling $T_c^{\rm
contact}\propto F^{1/3}$, from $\sigma\propto a\propto F^{1/3}$ in
Eq.~\eqref{eq:dTcmin}, can be experimentally tested by controlling/measuring the load under a point contact.

\textbf{Class II} ($C^{\rm exp} \gg C_{\rm vol}$): The node-separation
mechanism requires the relevant band crossing to be ungapped.  Spin-orbit or
crystal-field gaps quench the singular response, so that $C_{\rm uni}$ is no
longer orders of magnitude larger than the volumetric sensitivity $C_{\rm
vol} \equiv K\,dT_c/dP$ extracted from hydrostatic pressure experiments.
For materials in this class, $C_{\rm vol}$ severely underestimates the true
uniaxial coupling, but an independent non-elastic mechanism that explains the
large observed $C^{\rm exp}$ is already supported by direct experimental
evidence. Therefore, $C^{\rm exp}$ here serves as a consistency check
on that mechanism.  In PdSb, a hard W tip raises $T_c$ from 1.32 to
$\sim$3-4\,K while a soft Au tip does not~\cite{PdSb2022}, directly
identifying a uniaxial-stress-driven Lifshitz transition ($C^{\rm
exp}\sim90$\,K) that is not favored by hydrostatic pressure.  In Sr$_2$RuO$_4$,
$\Delta T_c^{\rm th}\simeq0.03$-0.12\,K falls far short of the $\sim$2\,K
enhancement, consistent with the van Hove singularity under uniaxial stress
established by strain-cell experiments~\cite{Steppke2017Science}.  In Si and
Ge, bulk superconductivity requires structural transitions at $\sim$11 and
$\sim$10\,GPa~\cite{Howlader2021_review}, far beyond any accessible tip
pressure, so the contact-induced transition is local and structural, not a
smooth strain-driven shift.

\textbf{Class III} ($C^{\rm exp} \gg C_{\rm vol}$): These materials share
the same characteristic as Class~II, namely $C_{\rm vol}$ from hydrostatic
experiments significantly underestimates the uniaxial coupling. But, in this case no
independent experimental signature has identified the responsible mechanism yet.
The values of $C^{\rm exp}$ in Table~\ref{tab:TISC_summary}, ranging from
$\sim$18\,K to $\sim$1520\,K, are therefore the principal result for these
systems. These constitute an experimental determination of the uniaxial coupling scale in each case, and can be used as precise targets for
dedicated microscopic investigation.  We note two boundary cases that
illustrate how the framework qualitatively organises different situations.
ZrSiS is a nodal-line semimetal whose non-symmorphic symmetry protects the
nodal lines regardless of spin-orbit coupling~\cite{Schoop2016NatComm}, yet
a uniaxial strain experiment to 0.34\% finds no Fermi-surface
modification~\cite{LorenzZrSiS2024} and no bulk superconductivity to at
least 20\,GPa~\cite{VanGennepZrSiS2019}, giving $C_{\rm uni}\lesssim1$\,K;
the tip-induced $T_c\approx7.5$\,K therefore cannot be strain-driven in the
usual sense, and has been attributed to local DOS
enhancement~\cite{Aggarwal2019ZrSiS}.  In TaAs$_2$ and NbAs$_2$, the
nominally Dirac-like crossings are gapped by spin-orbit
coupling~\cite{Zhang2020PRB}, quenching the node-separation mechanism; the
required $C^{\rm exp}\sim135$-149\,K greatly exceeds $C_{\rm vol}$, and
the microscopic origin remains to be identified spectroscopically.  In all
other Class~III systems the large gap between $C_{\rm th}$ and $C^{\rm exp}$
suggests that the dominant coupling channel is not captured by hydrostatic
pressure data alone; the tabulated $C^{\rm exp}$ values estimated using
independent bulk moduli and pressure-phase
data~\cite{WCbulkmod,ZhangCe2023,EichlerGey1972,AllenMartin1982} are intended to
guide a microscopic understanding in each case.

The treatment described here can also be used to predict TISC in previously untested materials.
We chose elemental Sb and Y as experimental tests because they represent two
physically distinct routes to satisfying the nucleation criterion.  Sb is
isostructural with grey As and, like it, hosts spin-polarised topological surface
states on the (111) face confirmed by ARPES and STS~\cite{SbTSSref}; these
states provide precisely the high-DOS coupling channel that makes $C^{\rm
exp}\sim66$\,K in grey As, and by the same reasoning one expects a comparable
$C^{\rm exp}$ in Sb.  With $K\simeq42$\,GPa, the framework then predicts
$T_c\simeq1$-5\,K. This is a window entirely inaccessible to
hydrostatic pressure, which requires a structural transition above
8.5\,GPa~\cite{SbPressureref,Hsieh2008} and gives $\Delta T_c^{\rm th}\ll1$\,K.
Y offers a contrasting case. It shares the hcp structure of Zr
and Ti, both Class~III hosts, and its Fermi level sits on the shoulder of a
broad $d$-band DOS peak that is highly sensitive to changes in the $c/a$ ratio.
Hydrostatic compression preserves $c/a$ to leading order and gives only
$C_{\rm vol}\lesssim4$\,K (bulk superconductivity requires $\gtrsim11$\,GPa~\cite{YunderP,YPressureref,YBulkMod}),
whereas uniaxial strain along the hexagonal axis directly distorts $c/a$ and
is expected to couple far more strongly. This is the same nature of enhancement seen in Zr and
Ti, but more pronounced given Y's softer bulk modulus and more steeply peaked DOS.
We formed Pt-Ir and Ag  point-contacts on both materials and observed TISC in
each~\cite{OurPRBref} with $T_c\approx2.8$\,K in Sb, and $T_c$ spanning
$\sim1.5$ to $\sim12$\,K across junctions in Y.  The wide spread of $T_c$ in Y essentially reveals the junction-to-junction variation in contact force, possibly following the load-scaling discussed in our framework.
Inverting the observed $T_c$ values gives $C^{\rm exp}\sim56$\,K for Sb  which is
consistent with grey As, and $C^{\rm exp}\approx160$-600\,K for Y, which
places Y decisively above Zr ($\approx113$\,K) and Ti ($\approx227$\,K) at the
top of the hcp-metal series.  Both materials fall in Class~III.

In conclusion, combining Hertzian contact mechanics with a
Ginzburg-Landau variational analysis we find a nucleation criterion
Eq.~\eqref{eq:dTcmin} for TISC/TESC.  Where the uniaxial strain-sensitivity is
independently known, as in Class~I semimetals with ungapped band
crossings, the criterion provides an explanation of TISC/TESC in those materials.
For all other materials, the criterion determines the uniaxial
coupling scale $C^{\rm exp}$ in each system. These values, ranging from $\sim$18\,K to
$\sim$1520\,K, can be taken as expeimental inputs for future microscopic theories.
The prediction and subsequent experimental confirmation of TISC in both Sb and Y demonstrate the predictive scope of the work across two distinct physical mechanisms.  The $T_c\propto F^{1/3}$ load scaling
is proposed as a universal test which can be confirmed by experiments where point-contact load can be measured/controlled. In addition to the 
conventional point contact geometries, Eq.~\eqref{eq:dTcmin} applies
to any confined stress field, making it a general diagnostic for
stress-nucleated order near dislocations, grain boundaries, and
patterned substrates.

We thank Dr. Debmalya Chakraborty for critically reading the manuscript and for sharing his valuable comments.

\end{document}